# Non-Hermitian edge burst of sound

*Hong-Yu Zou, Bing-Bing Wang, Yong Ge, Ke-Qi Zhao, Yu-Qi Chen, Hong-Xiang Sun,[*] Shou-Qi Yuan,[*] Haoran Xue,[*] and Baile Zhang[*]*

H.-Y. Zou, B.-B. Wang, Y. Ge, K.-Q. Zhao, Y.-Q. Chen, H.-X. Sun, S.-Q. Yuan

Research Center of Fluid Machinery Engineering and Technology, School of Physics and Electronic Engineering, Jiangsu University, Zhenjiang 212013, China.
E-mail: jsdxshx@ujs.edu.cn (H.-X. S.), shouqiy@ujs.edu.cn (S.-Q. Y.)

B.-B. Wang, H. R. Xue

Department of Physics, The Chinese University of Hong Kong, Shatin, Hong Kong SAR, China.
E-mail: haoranxue@cuhk.edu.hk (H. R. X.)

B. L. Zhang

Division of Physics and Applied Physics, School of Physical and Mathematical Sciences, Nanyang Technological University, 21 Nanyang Link, Singapore 637371, Singapore.
E-mail: blzhang@ntu.edu.sg (B. L. Z.)

B. L. Zhang

Centre for Disruptive Photonic Technologies, Nanyang Technological University, 21 Nanyang Link, Singapore 637371, Singapore.

Non-Hermitian band topology can give rise to phenomena with no counterparts in Hermitian systems. A well-known example is the non-Hermitian skin effect (NHSE), where Bloch eigenstates localize at a boundary, induced by a nontrivial spectrum winding number. In contrast, recent studies on lossy non-Hermitian lattices have uncovered an unexpected boundary-localized loss probability—a phenomenon that requires not only non-Hermitian band topology but also the closure of the imaginary (dissipative) gap. Here, we demonstrate the non-Hermitian edge burst in a classical-wave metamaterial: a lossy nonreciprocal acoustic crystal. We show that, when the imaginary gap remains closed, edge bursts can occur at the right boundary, left boundary, or both boundaries simultaneously, all under the same non-Hermitian band topology; the latter scenario is known as a bipolar edge burst. The occurrence of each scenario depends on the number and location of the imaginary gap closure points in the eigenenergy spectra. These findings generalize the concept of edge burst from quantum to classical wave systems, establish it as an intrinsic material property, and enrich the physics of the complex interplay between non-Hermitian band topology and other physical properties in non-Hermitian systems.

# 1. Introduction

Recent years have seen remarkable growth in the study of non-Hermitian physics,[1-3] particularly with the establishment of non-Hermitian band topology, which extends beyond the conventional framework of topological classification of matter. A paradigmatic example is the discovery of the non-Hermitian skin effect (NHSE),[4-36] where a large number of bulk eigenstates can localize at a boundary, either left or right, that is uniquely determined by the winding direction of eigenenergy spectrum in the complex energy plane. When the winding topology features a twist, the bulk eigenstates exhibit bipolar localization, accompanied by a Bloch-point-mediated extended state that interpolates between left- and right-localized eigenstates—a phenomenon referred to as the bipolar NHSE[20,21] (The detailed relationship between twisted-winding topology and bipolar NHSE is discussed in the Supporting Information). These NHSE phenomena have been experimentally observed across various physical platforms,[11,12,15-17,21-24,26-28,30,31,34,35] manifesting the universality of non-Hermitian band topology in all non-Hermitian systems.

In addition to the NHSE, recent studies have uncovered another boundary-induced phenomenon known as the non-Hermitian edge burst.[37-44] Specifically, in lossy systems, particles or wave packet energy can leak out (or be absorbed) at the lossy sites, resulting in a decrease in their probability of particle existence or wave packet energy within the system; such a decrease is referred to as the "loss probability". In this effect of non-Hermitian edge burst, particles or wave packets initially positioned far from the system's boundary display a loss probability that unexpectedly peaks at the boundary, with the peak's relative intensity increasing as the initial excitation moves further from the boundary. This phenomenon has recently been observed experimentally in photonic quantum walks that simulate a synthetic non-Hermitian periodic lattice.[43,44] Nevertheless, edge burst should be a universal wave phenomenon across quantum and classical domains and manifest as an intrinsic material property—yet neither aspect has been established in previous studies.

The current understanding of edge burst is that it is determined by two factors: (1) the NHSE, as fully governed by the non-Hermitian band topology, and (2) the closure of the gap—referred to as the imaginary gap—between the real axis and the imaginary part of the eigenenergy spectrum under the periodic boundary condition (PBC). Let us consider the simplest one-loop winding topology in the complex energy plane for a 1D lattice, as illustrated schematically in **Figure 1**a. An input wave launched from a bulk site propagates towards a boundary (the left boundary in the illustration) due to the NHSE. If the imaginary gap is open, the loss probability remains localized around the input site. However, with a closed imaginary gap, as shown in Figure 1b, a peak in loss probability appears at the left boundary. This behavior is well understood from recent theoretical[38] and experimental studies.[43,44]

In this work, we experimentally demonstrate multiple types of non-Hermitian edge bursts in a classical-wave metamaterial, realized via an acoustic crystal platform equipped with unidirectional amplifiers to implement nonreciprocal couplings.[21,35] This endeavor not only confirms the universality of edge burst as a wave phenomenon applicable to both quantum[37-44] and classical regimes, but also marks a significant advance by establishing edge burst as an intrinsic material property, rather than an effective simulation in the synthetic space.[43,44] We first realize both closed and open imaginary gaps in eigenfrequency spectra exhibiting a single-loop winding topology (similar to Figure 1a,b) to demonstrate the acoustic edge burst and its distinction from the previously demonstrated acoustic NHSE.[21,22,27,28,31] Unlike previous experiments,[43,44] which were limited to single-loop winding, we extend our study by constructing a twisted-winding topology. Such a twisted-winding topology guarantees bipolar NHSE, which, under the condition of imaginary gap closure, should lead to edge bursts occurring at both the left and right boundaries, known as bipolar edge bursts (Figure 1c).[38] Furthermore, our work uncovers a novel insight into the mechanism of edge bursts. By simultaneously utilizing nearest-neighbor and next-nearest-neighbor couplings in the 1D acoustic crystal, we can switch the bipolar edge burst to the previously observed edge burst at either the left or right boundary (Figure 1d,e), even when the imaginary gap remains closed and the non-Hermitian band topology remains unchanged. The different scenarios of edge burst depend on the number and location of imaginary gap closure points in the frequency spectra, which has not been previously investigated.

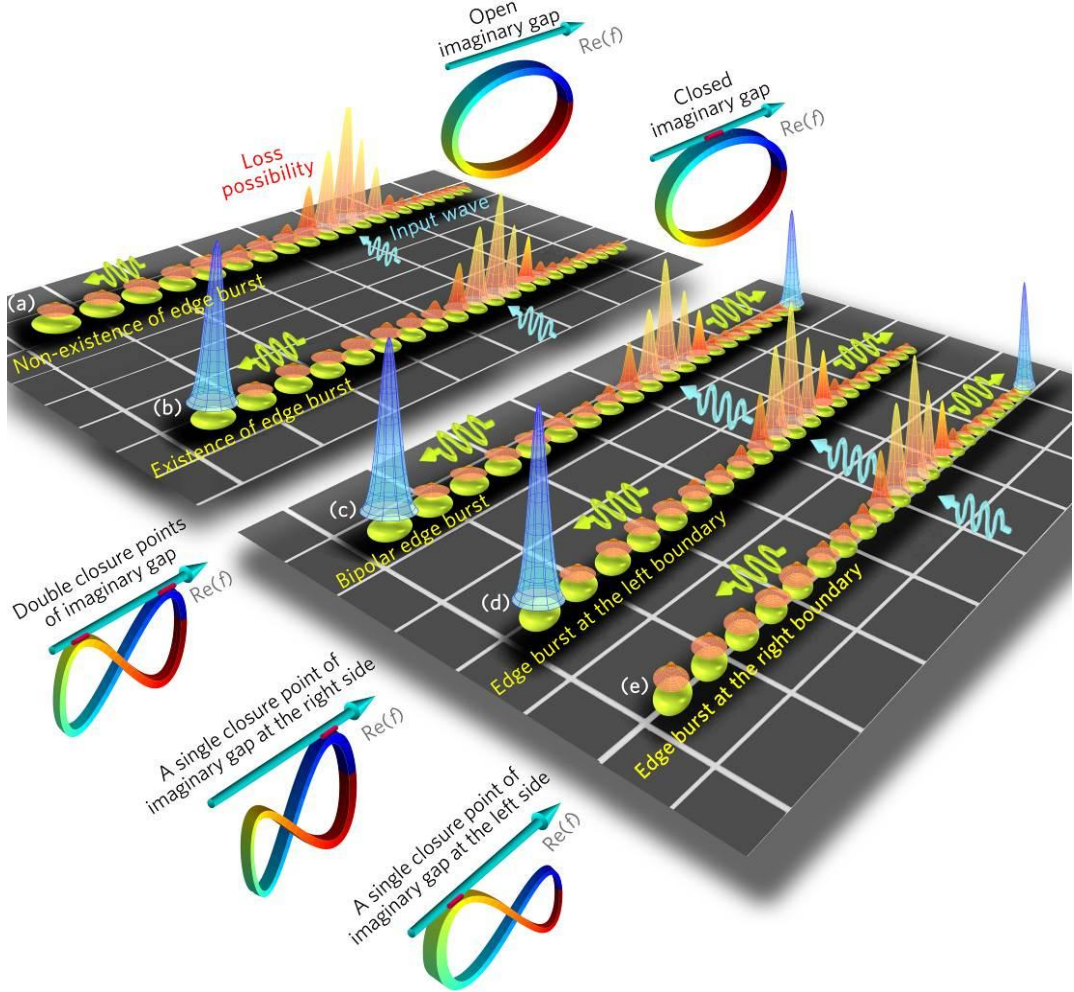

**Figure 1.** Schematic of loss probabilities in a 1D lossy system with different winding topologies. a) Non-existence and b) existence of edge burst for a single-loop winding topology with open and closed imaginary gaps, respectively. c) Bipolar edge burst for a twisted-winding topology with double closure points of the imaginary gap. Edge burst at only d) the left boundary or e) the right boundary when there is only one closure point of the imaginary gap. The right upper insets of (a) and (b) and left lower insets of (c)-(e) are the corresponding eigenfrequency spectra under the PBC. The color of the spectra represents the winding direction.

## 2. Edge Burst in 1D Acoustic Crystals with Single-Winding Topology

Firstly, we design a 1D acoustic crystal composed of $N=21$ resonators with the resonance frequency $f_0$ and the intrinsic onsite loss $\gamma$. The adjacent resonators are coupled with a pair of narrow tubes, and thus leading to a reciprocal coupling $\kappa$ (The detailed structure parameters of the acoustic crystal are presented in the Experimental Section). By fitting the measured dispersion relationship and the pressure amplitude profile along the acoustic crystal with the theoretical ones calculated by the Green's function,[45] we can obtain the values of $\kappa$, $\gamma$ and $f_0$ as -22.5, 8 and 2180 Hz (The detailed experimental set-up and fitting procession are provided in the Supporting Information). In addition, we introduce a unidirectional amplifier into a basic unit cell of the acoustic crystal to realize nonreciprocal coupling $\kappa_a$, in which the value and sign of $\kappa_a$ can be effectively changed by tuning the gain factor and the connection path of the amplifier[21,35,46] (see the Supporting Information).

To explore the interplay between the NHSE and the closure of the imaginary gap, and realize the edge burst phenomenon in the classical wave material structure, we then design a 1D nonreciprocal acoustic crystal with a nearest-neighbor positive coupling $\kappa_{a,1}$ (shown in **Figure 2**a), in which the adjacent resonators are connected with a unidirectional amplifier. To obviously display the connection path of the amplifier, we present the enlarged view of open rectangle I in the right inset. Figure 2b shows the tight-binding (TB) model of the 1D nonreciprocal acoustic crystal, and the corresponding complex eigenfrequency spectrum under the PBC can be written as:

$$f_k = \kappa_{a,1} e^{-ika} + 2\kappa \cos ka + f_0 - i\gamma , \qquad (1)$$

where $a$ is the lattice constant and $k$ is the Bloch wave vector.

As shown in Figure 2c,d, we select the values of $\kappa_{a,1}$ as 8 (=$\gamma$) and 4 Hz (<$\gamma$), respectively, and calculate the eigenfrequency spectra under both the PBC using Equation (1) and the open boundary condition (OBC). We can see that both spectra under the PBC wind along a loop with a single direction, and the imaginary gaps are closed and open for $\kappa_{a,1}$=8 and 4 Hz, respectively. Moreover, in both cases ($\kappa_{a,1}$=8 and 4 Hz), the eigenfrequency spectra under the OBC are enclosed by those under the PBC, confirming the occurence of NHSE.[13] To demonstrate this, we simulate the eigenstates under the OBC, and find that the eigenstates are located at the left boundaries ($n$=1). The amplitude distributions of the eigenstates are presented in the Supporting Information. Therefore, the TB model with $\kappa_{a,1}$=8 Hz satisfies the condition of the edge burst, however, that does not satisfy for $\kappa_{a,1}$=4 Hz.

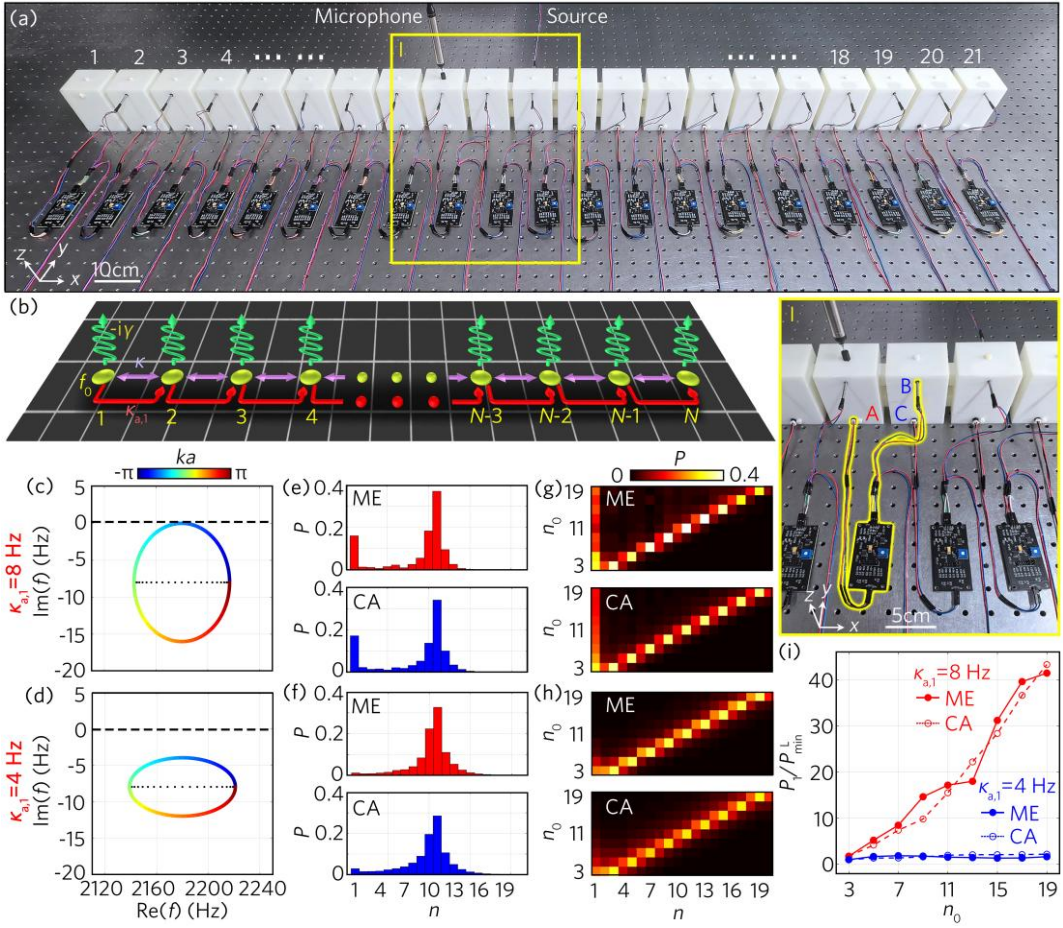

**Figure 2.** Observation of edge burst in a 1D acoustic crystal with nearest-neighbor nonreciprocal positive coupling. a) Photograph of the 1D nonreciprocal acoustic crystal composed of 21 resonators, and the adjacent resonators are connected with an amplifier. The right inset is an enlarged view in open rectangle I. b) Schematic of the TB model of the 1D nonreciprocal acoustic crystal. Calculated eigenfrequency spectra under the PBC (colored points) and OBC (black points) in the complex frequency plane for c) $\kappa_{a,1}=$ 8 and d) 4 Hz. Measured (ME) and calculated (CA) loss probabilities $P_n$ with $n_0=11$ for e) $\kappa_{a,1}=$ 8 and f) 4 Hz, in which the red and blue bars are the measured and calculated results, respectively. Measured and calculated $P_n$ with $n_0=$3, 5,...,17 and 19 for g) $\kappa_{a,1}=$ 8 and h) 4 Hz. i) Measured and calculated $P_1/P_{min}^L$ with $n_0=$3, 5,...,17 and 19 for $\kappa_{a,1}=$8 (the red points and lines) and 4 Hz (the blue points and lines).

To describe the edge burst phenomenon in the acoustic crystal of Figure 2a, we theoretically derive the loss probability $P_n$ at the $n$-th site of the corresponding TB model (see the Supporting Information), which can be expressed as:

$$P_n = \Delta f \sum_{f=f_{min}}^{f_{max}} \begin{bmatrix} p_n(f) \\ p_{n+1}(f) \end{bmatrix}^\dagger \begin{bmatrix} \gamma(1+\delta_{n,1}) & \dfrac{\kappa_{a,1}^*}{i} \\ i\kappa_{a,1} & \gamma(1+\delta_{n,N-1}) \end{bmatrix} \begin{bmatrix} p_n(f) \\ p_{n+1}(f) \end{bmatrix}, \qquad (2)$$

where $\delta_{n,1}$ and $\delta_{n,N-1}$ is the Kronecker delta symbol, the maximum frequency $f_{max}=$2400 Hz, the minimum frequency $f_{\min}$=1900 Hz, the frequency interval $\Delta f$=1 Hz, and $p_n(f)$ is the pressure spectrum in the $n$-th resonator. Here, when the gain factors of all the amplifiers are precisely tuned to attain $\kappa_{a,1}$=8 and 4 Hz separately, we select the source position $n_0$=11, and measure $p_n(f)$ in each resonator. The measured amplitude spectra $|p_n(f)|$ are presented in the Supporting Information. By substituting the measured spectra $p_n(f)$ into Equation (2), we can obtain the loss probability $P_n$ for $\kappa_{a,1}$=8 and 4 Hz, which are presented as the red bars in Figure 2e,f, respectively. For both cases, we observe that the values of $P_n$ decay from the $n_0$-th site, and the distributions are asymmetric around the $n_0$-th site, which is attributed to the NHSE. In addition, there exist some differences for both cases. For $\kappa_{a,1}$=8 Hz (Figure 2e), an obvious peak of $P_n$ exists at the left boundary ($n$=1), showing a typical characteristic of the edge burst. However, for $\kappa_{a,1}$=4 Hz (Figure 2f), the value of $P_n$ is almost zero at the left boundary, indicating that the edge burst does not exist when the imagery gap is open. Furthermore, we theoretically calculate the pressure spectra $p_n(f)$ for $\kappa_{a,1}$=8 and 4 Hz based on $p_n(f)=\langle n|\hat{G}(f)|n_0\rangle$, where $\hat{G}(f)=(f\hat{I}-\hat{H})^{-1}$ is the Green's function,[45] and $\hat{I}$ is the identity matrix. The theoretical results of $|p_n(f)|$ are presented in the Supporting Information. By substituting the theoretical values of $p_n(f)$ into Equation (2), we theoretically calculate the loss probability $P_n$ for both cases (the blue bars in Figure 2e,f), which agree well with the measured ones.

Next, we discuss the influence of the source position $n_0$ on the loss probability $P_n$, we here select $n_0$=3, 5, ...,17 and 19, and measure and calculate the corresponding values of $P_n$, which are shown in Figure 2g,h, respectively. We observe that, by adjusting the source position, the edge burst always exists for $\kappa_{a,1}$=8 Hz, but does not exist for $\kappa_{a,1}$=4 Hz, indicating the robustness of the edge burst for the source position. In addition, to quantitatively characterize the edge burst with different values of $n_0$, we define the relative value of $P_n$ for the peak at the left boundary as $P_1/P_{\min}^L$, where $P_{\min}^L=\min\{P_1,P_2,...,P_{n_0-1}\}$. Based on the results in Figure 2g,h, we obtain the relationship between $P_1/P_{\min}^L$ and $n_0$ for $\kappa_{a,1}$=8 and 4 Hz, which is shown in Figure 2i. It is observed that the value of $P_1/P_{\min}^L$ gradually increases when the sound source moves away from the left boundary ($n$=1) for $\kappa_{a,1}$=8 Hz, while that remains unchanged for $\kappa_{a,1}$=4 Hz.

The measured results agree well with the calculated ones.

To explain it, we theoretically derive the algebraic forms of $P_n$ for the closed ($|\kappa_{a,1}|=\gamma$) and the open ($|\kappa_{a,1}|<\gamma$) imaginary gaps based on the TB model in Figure 2b. We can see that the existence and non-existence of the edge burst arise from the powered and exponential terms of $P_n$, respectively (see the Supporting Information). Moreover, in Figure 2f, a slight local peak in loss probability is observed at the 1st site in both the calculated and measured results. The reason for this is explained in the Supporting Information. In addition, we also discuss the case of the eigenfrequency spectrum crossing the real axis when $\kappa_{a,1}>\gamma$, which is presented in the Supporting Information.

We also experimentally demonstrate the edge burst in the acoustic crystal with the nonreciprocal negative coupling ($\kappa_{a,1}$=-8 Hz), and the other parameters are the same as those in Figure 2. Compared with the result for $\kappa_{a,1}$=8 Hz in Figure 2e, the edge burst is observed at the right boundary of the acoustic crystal ($n$=$N$-1) for $\kappa_{a,1}$=-8Hz, which are presented in the Supporting Information. Furthermore, we study the influences of the onsite loss $\gamma$ on the edge burst. To enlarge the value of $\gamma$ of the acoustic crystal, we place sponges in a hole of each resonator, and then tune the gain factor of the amplifier to satisfy the condition of $|\kappa_{a,1}|=\gamma$. Compared with the result without the sponges, the peak of the edge burst becomes sharper with the sponges, and the relative value of $P_n$ at the peak changes greatly with the movement of the sound source. The measured and calculated results are presented in the Supporting Information.

## 3. Bipolar Edge Burst in 1D Acoustic Crystals with Twisted-Winding Topology

In addition to the aforementioned results, we experimentally study the bipolar edge burst in the 1D acoustic crystal with a next-nearest-neighbor nonreciprocal positive coupling $\kappa_{a,2}$ (shown in **Figure 3**a), in which the connecting path of the amplifier is presented in the enlarged view of open rectangle II. Figure 3b schematically shows the TB model, and the complex eigenfrequency spectrum under the PBC can be expressed as:

$$f_k = \kappa_{a,2} e^{-2ika} + 2\kappa \cos ka + f_0 - i\gamma, \tag{3}$$

where the next-nearest-neighbor nonreciprocal couplings are selected as $\kappa_{a,2}$=8 and 4 Hz. Figure 3c,d shows the theoretically calculated eigenfrequency spectra for $\kappa_{a,2}$=8 and 4 Hz based on Equation (3), respectively. We can see that, both eigenfrequency spectra under the PBC wind along two oppositely oriented loops in the complex frequency planes. The loops in the range Re(*f*)>2180 Hz wind in the clockwise direction, while those in the range Re(*f*)<2180 Hz wind at the opposite direction, showing the typical characteristics of a twisted-winding topology. Moreover, as shown in Figure 3c, both loops are tangent to the real axis when $\kappa_{a,2}$=8 Hz (=*γ*), indicating the characteristic of the closed imaginary gap. However, when the value of $\kappa_{a,2}$ is smaller than that of *γ* (Figure 3d), the imaginary gap is open. Moreover, in the ranges Re(*f*)>2180 Hz and Re(*f*)<2180 Hz, the eigenfrequency spectra under the OBC are enclosed by those under the PBC winding in the clockwise and anticlockwise directions, and the eigenstates under the OBC are located at the left (*n*=1) and right boundaries (*n*=N), respectively (see the Supporting Information), which indicates that the bipolar NHSE occurs. Therefore, the TB model with $\kappa_{a,2}$=8 Hz can meet the condition of the excitation of the bipolar edge burst, while that for $\kappa_{a,2}$=4 Hz does not.

To describe the phenomenon of bipolar edge burst in the acoustic crystal of Figure 3a, we rewrite the expression of $P_n$ as:

$$P_n = \Delta f \sum_{f=f_{min}}^{f_{max}} \begin{bmatrix} p_n(f) \\ p_{n+1}(f) \\ p_{n+2}(f) \end{bmatrix}^{\dagger} \begin{bmatrix} \frac{2\gamma}{3}D_n^{(1)} & 0 & \frac{\kappa_{a,2}^*}{i} \\ 0 & \frac{2\gamma}{3}D_n^{(2)} & 0 \\ i\kappa_{a,2} & 0 & \frac{2\gamma}{3}D_n^{(3)} \end{bmatrix} \begin{bmatrix} p_n(f) \\ p_{n+1}(f) \\ p_{n+2}(f) \end{bmatrix}, \qquad (4)$$

where $D_n^{(1)} = 2\delta_{n,1} + \delta_{n,2}/2 + 1$, $D_n^{(2)} = (\delta_{n,1} + \delta_{n,N-2})/2 + 1$ and $D_n^{(3)} = 2\delta_{n,N-2} + \delta_{n,N-3}/2 + 1$. The derivation is presented in the Supporting Information. Next, we measure $p_n(f)$ in each resonator for $n_0$=11 (see the Supporting Information). By substituting $p_n(f)$ into Equation (4), we can obtain the loss probability $P_n$ (the red bars in Figure 3e,f). As shown in Figure 3e, for $\kappa_{a,2}$=8 Hz, the value of $P_n$ decays gradually from the $n_0$-th site, and almost becomes zero when approaching both boundaries. However, there exist obvious peaks at both boundaries of the acoustic crystal, showing the typical characteristic of the bipolar edge burst (the red bars in Figure 3e). The peaks of $P_n$ at both boundaries stem from two different winding directions of the eigenfrequency spectrum at the closure points of the imaginary gap. For $\kappa_{a,2}$=4 Hz, the peak of $P_n$ does not exist at both boundaries (Figure 3f), revealing that the bipolar edge burst is non-existent when the imaginary gap is open. In addition, we theoretically calculate the values of $P_n$ for both cases by using the Green's function (the blue bars in Figure 3e,f), which match well with the measured ones.

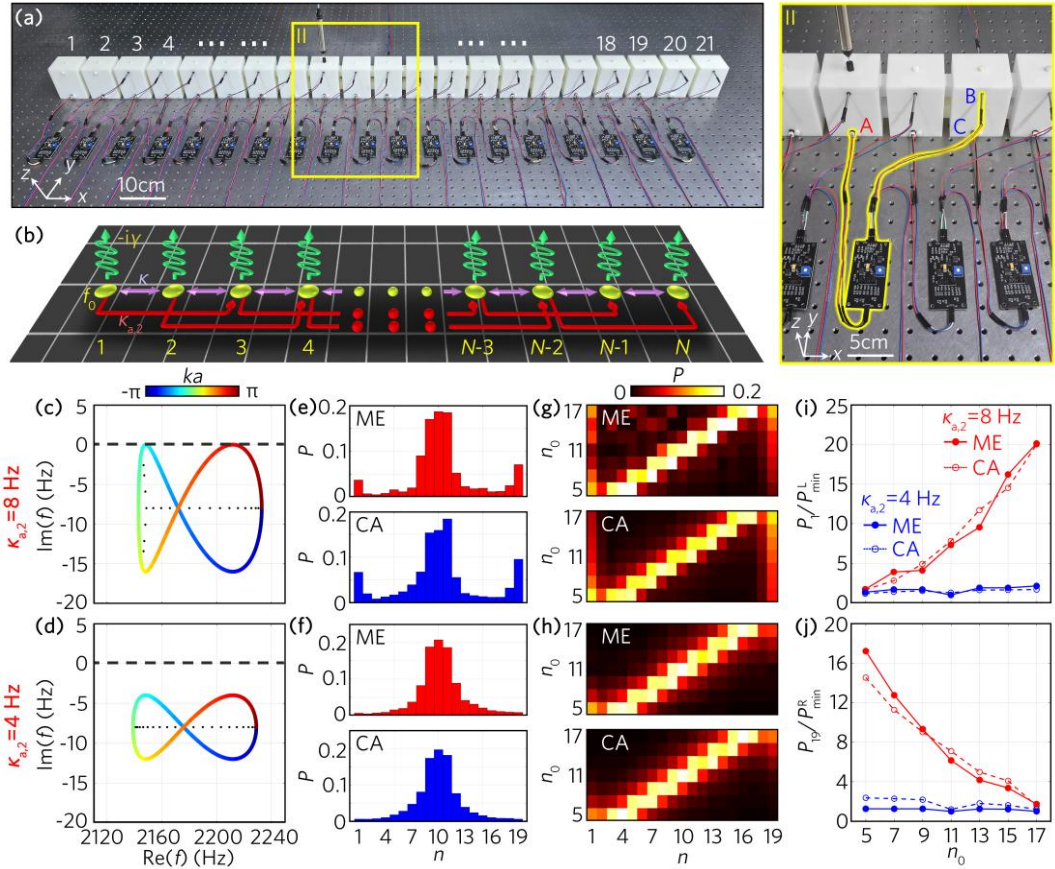

**Figure 3**. Observation of bipolar edge burst in the 1D acoustic crystal with a next-nearest-neighbor nonreciprocal positive coupling. a) Photograph of the 1D nonreciprocal acoustic crystal, and the next-nearest-neighbor resonators are connected with an amplifier. The right inset is an enlarged view in open rectangle II. b) Schematic of the TB model of the 1D nonreciprocal acoustic crystal. Calculated eigenfrequency spectra under the PBC (colored points) and OBC (black points) in the complex frequency plane for c) $\kappa_{a,2}$= 8 and d) 4 Hz. Measured (the red bars) and calculated (the blue bars) loss probabilities $P_n$ with $n_0$=11 for e) $\kappa_{a,2}$=8 and f) 4 Hz. Measured and calculated $P_n$ with $n_0$=5, 7,...,15 and 17 for g) $\kappa_{a,2}$=8 and h) 4 Hz. Measured and calculated i) $P_1/P_{min}^L$ and j) $P_{N-2}/P_{min}^R$ with $n_0$=5, 7,...,15 and 17 for $\kappa_{a,2}$=8 (the red points and lines) and 4 Hz (the blue points and lines).

Similarly, we measure and calculate the loss probability $P_n$ with $n_0$=5,7,...,15 and 17 for $\kappa_{a,2}$=8 and 4 Hz, which are shown in Figure 3g,h, respectively. It is observed that, for $\kappa_{a,2}$=8 Hz, the peaks at both boundaries always exist with different values of $n_0$ (Figure 3g), while those for $\kappa_{a,2}$=4 Hz do not exist (Figure 3h), indicating the robustness of the bipolar edge burst on the source position. To quantitatively characterize the bipolar edge burst with different values of $n_0$, we define the relative value of $P_n$ for the peak at the right boundary as $P_{N-2}/P_{\min}^{R}$, in which $P_{\min}^{R} = \min\{P_{n_0+1}, P_{n_0+2},...,P_{N-3}, P_{N-2}\}$. The definition of $P_{\min}^{L}$ at the left boundary is the same as that of the edge burst. As shown in Figure 3i,j, the values of $P_1/P_{\min}^{L}$ and $P_{N-2}/P_{\min}^{R}$ increase gradually by moving the source away from the left and right boundaries for $\kappa_{a,2}$=8 Hz (the red points and lines), respectively, while both parameters almost remain constant for $\kappa_{a,2}$=4 Hz (the blue points and lines). The measured results agree well with the calculated ones.

**4. Boundary-Selectable Edge Burst in 1D Acoustic Crystals with Twisted-Winding Topology**

Finally, we reveal a new insight that even when imaginary gap is closed and the topology of eigenfrequency spectrum remains unchanged, the edge burst can take different forms dependent on how the imaginary gap is closed. To demonstrate this, we introduce an additional nearest-neighbor nonreciprocal coupling $\kappa_{a,1}$ into the next-nearest-neighbor nonreciprocal acoustic crystal in Figure 3a. **Figure 4**a shows the photograph of a 1D nonreciprocal acoustic crystal with both $\kappa_{a,2}$ and $\kappa_{a,1}$, in which the next-nearest-neighbor and nearest-neighbor resonators on the front and rear sides are connected with 2 series of amplifiers with the nonreciprocal couplings of $\kappa_{a,2}$ and $\kappa_{a,1}$, respectively. The bottom insets are the enlarged forward and backward views in open rectangle III. Similarly, the positive and negative values of $\kappa_{a,1}$ can be realized by inserting the loudspeaker of the amplifier into holes B and C, respectively (see the 2 bottom insets on the right side). The corresponding TB model is shown in Figure 4b, and the complex eigenfrequency spectrum under the PBC can be rewritten as:

$$f_k = \kappa_{a,2} e^{-2ika} + \kappa_{a,1} e^{-ika} + 2\kappa \cos ka + f_0 - i\gamma, \tag{5}$$

where $\kappa_{a,1}=\pm 4$ Hz and $\kappa_{a,2}=5$ Hz. Based on Equation (5), we theoretically calculate the eigenfrequency spectra for $\kappa_{a,1}=\pm 4$ Hz, which are presented in Figure 4c,d, respectively. By comparing eigenfrequency spectra in Figure 4c,d with that in Figure 3c, we can see that, although the coupling $\kappa_{a,1}$ is additionally introduced, the spectra under the PBC in Figure 4c,d still have the same characteristics of the twisted-winding topology, and the imaginary gaps remain closed. Moreover, similar to Figure 3c,d, the spectra under the PBC with such winding topology still enclose those under the OBC, and the eigenstates under the OBC with different eigenfrequencies are still located at either left or right boundary, which are the characteristics of the bipolar NHSE (see the Supporting Information). However, the numbers and locations of the closure points of the imaginary gap are changed. For both cases in Figure 4c,d, there only exists a single closure point of the imaginary gap. The closure point of the imaginary gap exists in the loop on the left side of the spectrum for $\kappa_{a,1}=4$ Hz (Figure 4c), while that exists in the loop on the right side for $\kappa_{a,1}=-4$ Hz (Figure 4d).

Here, for the model in (Figure 4b), the expression of the loss probability $P_n$ can be extended as:

$$P_n = \Delta f \sum_{f=f_{\min}}^{f_{\max}} \begin{bmatrix} p_n(f) \\ p_{n+1}(f) \\ p_{n+2}(f) \end{bmatrix}^{\dagger} \begin{bmatrix} \frac{2\gamma}{3}D_n^{(1)} & \frac{\kappa_{a,1}^*}{2i}(1+\delta_{n,1}) & \frac{\kappa_{a,2}^*}{i} \\ \frac{i\kappa_{a,1}}{2}(1+\delta_{n,1}) & \frac{2\gamma}{3}D_n^{(2)} & \frac{\kappa_{a,1}^*}{2i}(1+\delta_{n,N-2}) \\ i\kappa_{a,2} & \frac{i\kappa_{a,1}}{2}(1+\delta_{n,N-2}) & \frac{2\gamma}{3}D_n^{(3)} \end{bmatrix} \begin{bmatrix} p_n(f) \\ p_{n+1}(f) \\ p_{n+2}(f) \end{bmatrix}, \quad (6)$$

whose derivation is presented in the Supporting Information. Similarly, we measure and calculate $p_n(f)$ for $\kappa_{a,1}=\pm 4$ Hz in each resonator for $n_0=11$ (see the Supporting Information). By substituting $p_n(f)$ into Equation (6), we can obtain the loss probability $P_n$ for both cases, which are shown in Figure 4e,f. Both the measured and calculated results show that, by selecting $\kappa_{a,1}=4$ Hz, there only exists a single peak of $P_n$ located at the left boundary, which can be switched to the right boundary of the acoustic crystal when selecting $\kappa_{a,1}=-4$ Hz.

Figure 4g,h presents the measured and calculated $P_n$ with $n_0=5, 7,...,15$ and 17,

respectively. It is observed that, for different values of $n_0$, the peak of $P_n$ always exists at the left and right boundaries for $\kappa_{a,1}$=4 and -4 Hz, respectively, indicating the robustness of the switch of both types of edge bursts on $n_0$. Finally, based on the results in Figure 4g,h, we obtain the values of $P_1/P_{\min}^L$ and $P_{N-2}/P_{\min}^R$. As shown in Figure 4i,j, by moving the position of the source to the right boundary, the value of $P_1/P_{\min}^L$ increases gradually for $\kappa_{a,1}$=4 Hz, while that is almost unchanged for $\kappa_{a,1}$=-4 Hz. However, the value of $P_{N-2}/P_{\min}^R$ remains constant for $\kappa_{a,1}$=4 Hz, and gradually reduces for $\kappa_{a,1}$=-4 Hz.

Therefore, despite the closed imaginary gap and the same twisted-winding topology of eigenfrequency spectra under the PBC, the phenomena of edge burst can be perfectly distinct: the bipolar edge burst (Figure 3e), the edge burst at either the left (Figure 4e) or right boundary (Figure 4f), which is due to the different numbers and locations of imaginary gap closure points in the frequency spectra.

## 5. Conclusion

In conclusion, we have experimentally demonstrated multiple types of edge bursts in a 1D acoustic crystal featuring both nearest-neighbor and next-nearest-neighbor nonreciprocal couplings. Our work not only extends the edge burst phenomenon from quantum systems[37-44] to classical wave platforms, but also establishes it as an intrinsic material property. Moreover, the number and location of the imaginary gap closure points can serve as tuning parameters to manipulate the edge burst phenomenon, such as switching the peak of loss probability from one boundary to another, even when the non-Hermitian topology and imaginary gap closure remain unchanged. It is noteworthy that while our implementation employs nearest- and next-nearest-neighbor couplings to realize edge burst phenomena, these specific design elements are not essential requirements. Alternative approaches, including loss-gain systems[29] and combinations of loss, artificial gauge fields, and Floquet engineering,[34] may facilitate the realization of edge bursts. Furthermore, edge burst phenomena are not exclusive to acoustic systems, but may also be observed across various non-Hermitian platforms, such as photonic systems[17] and electrical circuits.[24] Most significantly, these phenomena can emerge even beyond the framework of TB model.[28,33,36] Our work significantly broadens the scope of edge burst and enriches the underlying physics of edge burst, which is a substantial conceptual and technical advance.

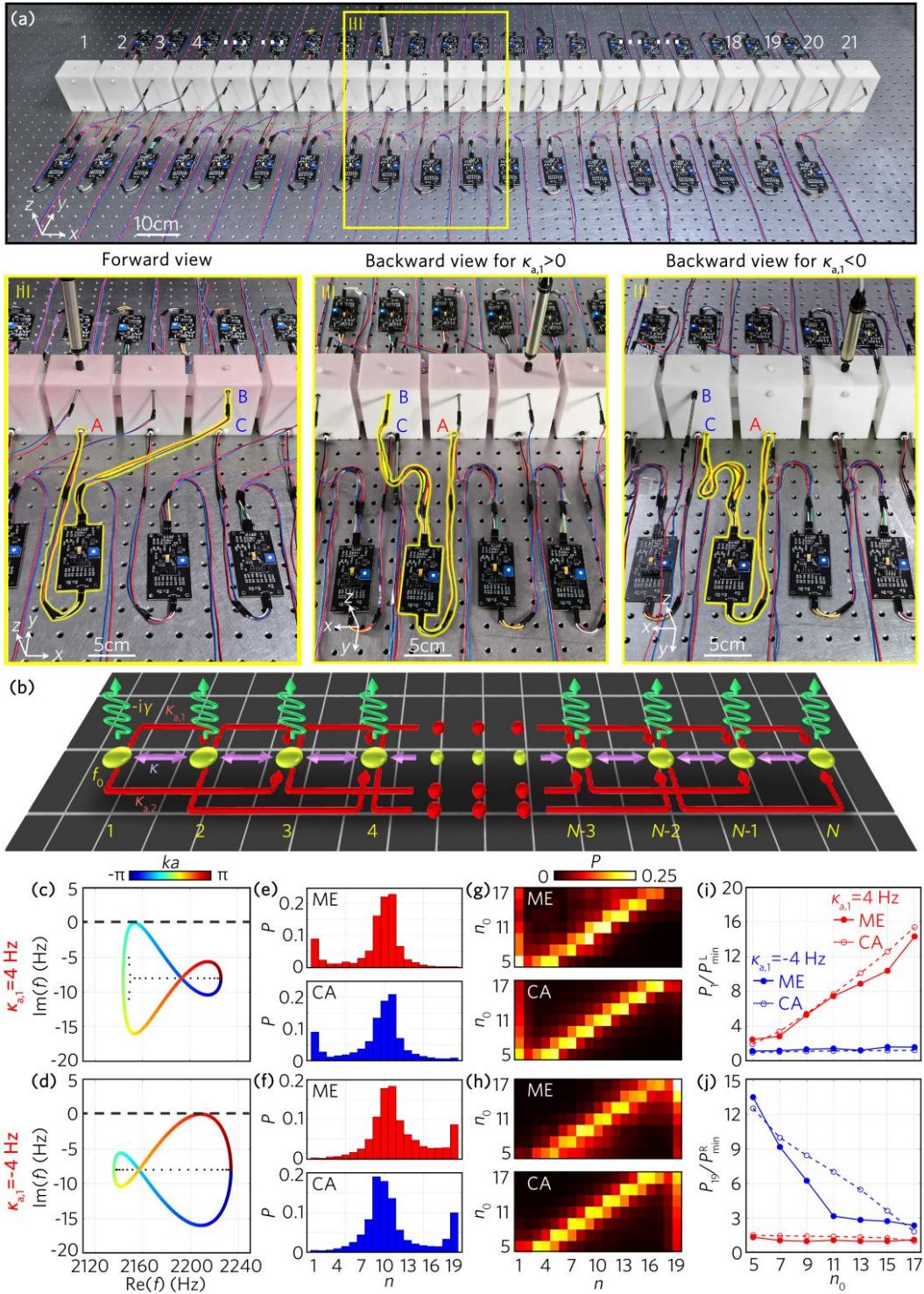

**Figure 4.** Observation of edge burst at either edge of the 1D acoustic crystal with both nearest-neighbor and next-nearest-neighbor nonreciprocal couplings. a) Photograph of the 1D nonreciprocal acoustic crystal, and the next-nearest-neighbor and nearest-neighbor resonators on the front and rear sides are connected with two series of amplifiers to realize the couplings $\kappa_{a,2}$ and $\kappa_{a,1}$, respectively. The bottom insets are the enlarged forward and backward views of open rectangle III. The positive and negative values of $\kappa_{a,1}$ can be achieved by inserting the loudspeaker of the amplifier into holes B and C, respectively (the two bottom insets at the right side). b) Schematic of the TB model of the 1D nonreciprocal acoustic crystal. Calculated eigenfrequency spectra under the PBC (colored points) and OBC (black points) in the complex frequency plane for c) $\kappa_{a,1}=4$ and d) -4 Hz. Measured (the red bars) and calculated (the blue bars) loss probabilities $P_n$ with $n_0=11$ for e) $\kappa_{a,1}=4$ and f) -4 Hz. Measured and calculated $P_n$ with $n_0=5,7,...,15$ and 17 for g) $\kappa_{a,1}=4$ and h) -4 Hz. Measured and calculated i) $P_1/P_{\min}^{L}$ and j) $P_{N-2}/P_{\min}^{R}$ with $n_0=5, 7, ...,15$ and 17 for $\kappa_{a,1}=4$ (the red points and lines) and -4 Hz (the blue points and lines).

**Experimental Section**

*Sample Fabrication:* In the experiment, all samples are fabricated by the 3D printing technique with epoxy resin. The sample of 1D acoustic crystal consists of 21 basis unit cells composed of two resonators coupled with a pair of narrow tubes, which is shown in **Figure S13** (see the Supporting Information). The size of each resonator are $l_x=7$ cm, $l_y=6$ cm and $l_z=9$ cm. and that of both tubes $w_1=1.8$ cm and $w_2=1$ cm, and the distance between the two tubes $w_3=4$ cm. The wall thickness $d=0.5$ cm. The lattice constant of the 1D acoustic crystal is $a=l_x+w_1=8.8$ cm. To insert the microphone and loudspeaker of the amplifier, we drill small and big holes on both front and back sides of each resonator. The radii of the small and big holes are 0.2 cm and 0.35 cm, respectively. Moreover, to insert the sound source and the microphone, we design the holes with the radius of 0.2 cm on the top surfaces of each resonator.

*Experimental Set-up and Statistical Analysis:* We use a power amplifier to drive a balanced armature speaker with a radius of 1 mm and launch a broadband sound signal within the range of 1900-2400 Hz (the frequency resolution is 1 Hz), which is guided into the sample through a narrow tube with the radius of 1.5 mm. In the experiments, to measure the sound pressure profile $p_n(f)$ along the 1D acoustic crystal, we insert a sound source into the backside of the $n_0$-th resonator. The sound signal is detected by the microphone (B&K type-4182) inserted into the hole on the top surface of each resonator. The amplitude and phase of the signal can be recorded by the software PULSE Labshop (B&K Type 3160-A-022).

**Supporting Information**

Supporting Information is available from the Wiley Online Library or from the author.

**Acknowledgements**

H.-Y. Zou and B.-B. Wang contributed equally to this work. This work is supported by the National Natural Science Foundation of China under Grants No. 12274183 and No.12174159, the National Key Research and Development Program of China under Grant No. 2020YFC1512403, the Jiangsu Qing Lan Project, the Postgraduate Research and Practice Innovation Program of Jiangsu Province under Grant No. KYCX23_3746, the start-up fund and the direct grant of the Chinese University of Hong Kong under Grant No. 4053675, the Singapore National Research Fundation Competitive Research Program under Grant No. NRF-CRP23-2019-0007, and the Singapore Ministry of Education Academic Research Fund Tier 2 under Grant No. MOE-T2EP50123-0007 and Tier 1 under Grant No. RG139/22 and No. RG81/23.

**Conflict of Interest**

The authors declare no conflict of interest.

**Keywords**

Non-Hermitian edge burst, acoustic crystal, nonreciprocal coupling, non-Hermitian skin effect, imaginary gap closure